\documentclass[12pt]{article}
\usepackage{psfig}
\def\toprule{\hline\hline}
\def\colrule{\hline}
\def\botrule{\hline\hline}
\def\tbl{\caption}
\def\refcite{\cite}
\RequirePackage{xspace}
\usepackage{relsize}
\input rotate.tex

\def\etal{{\em et~al.}}
\def\eg{{\em e.g.}}

\def\thetaH{\ensuremath{\theta\kern-0.1em_H}}

\def\BF{$B$ Factory}
\def\invfb   {\ensuremath{\mbox{\,fb}^{-1}}\xspace}
\def\DeltaE  {\mbox{$\Delta E$}\xspace}
\def\deltae  {\ensuremath{\DeltaE^*}\xspace}
\def\mes     {\mbox{$m_{\rm ES}$}\xspace}
\def\BR      {{\ensuremath{\cal B}\xspace}}

\def\Bu      {\ensuremath{B^+}\xspace}
\def\Bbar    {\kern 0.18em\overline{\kern -0.18em B}{}\xspace}
\def\B       {\ensuremath{B}\xspace}
\def\Bz      {\ensuremath{B^0}\xspace}
\def\BB      {\ensuremath{B\Bbar}\xspace}
\def\Bp      {\ensuremath{\Bu}\xspace}
\def\Bm      {\ensuremath{\Bub}\xspace}
\def\Bzb     {\ensuremath{\Bbar^0}\xspace}
\def\Bub     {\ensuremath{B^-}\xspace}

\def\Kstarz  {\ensuremath{K^{*0}}\xspace}

\def\Kstar   {\ensuremath{K^*}\xspace}
\def\Kstarzb {\ensuremath{\Kbar^{*0}}\xspace}
\def\Kstarp  {\ensuremath{K^{*+}}\xspace}
\def\Km      {\ensuremath{K^-}\xspace}
\def\KS      {\ensuremath{K^0_{\scriptscriptstyle S}}\xspace}
\def\kaon    {\ensuremath{K}\xspace}
\def\Kstarm  {\ensuremath{K^{*-}}\xspace}
\def\Kbar    {\kern 0.2em\overline{\kern -0.2em K}{}\xspace}
\def\Kp      {\ensuremath{K^+}\xspace}

\def\pip     {\ensuremath{\pi^+}\xspace}
\def\piz     {\ensuremath{\pi^0}\xspace}
\def\pim     {\ensuremath{\pi^-}\xspace}

\def\g       {\ensuremath{\gamma}\xspace}
\def\en      {\ensuremath{e^-}\xspace}  
\def\ep      {\ensuremath{e^+}\xspace}
\def\qqbar   {\ensuremath{q\overline q}\xspace}

\def\babar{\mbox{\slshape B\kern-0.1em{\footnotesize A}\kern-0.1em
    B\kern-0.1em{\footnotesize A\kern-0.2em R}}}
\def\BaBar{\babar\xspace}
\def\referencebabar{\mbox{\slshape B\kern-0.1em{\footnotesize A}\kern-0.1em
    B\kern-0.1em{\footnotesize A\kern-0.2em R}}\xspace}
\def\scriptbabar{\mbox{\slshape B\kern-0.1em{\smaller A}\kern-0.1em
    B\kern-0.1em{\smaller A\kern-0.2em R}}\xspace}

\newcommand{\gev}{\ensuremath{\mathrm{\,Ge\kern -0.1em V}}\xspace}
\newcommand{\mev}{\ensuremath{\mathrm{\,Me\kern -0.1em V}}\xspace}
\newcommand{\gevc}{\ensuremath{{\mathrm{\,Ge\kern -0.1em V\!/}c}}\xspace}
\newcommand{\mevc}{\ensuremath{{\mathrm{\,Me\kern -0.1em V\!/}c}}\xspace}
\newcommand{\gevcc}{\ensuremath{{\mathrm{\,Ge\kern -0.1em V\!/}c^2}}\xspace}
\newcommand{\mevcc}{\ensuremath{{\mathrm{\,Me\kern -0.1em V\!/}c^2}}\xspace}

\newsavebox{\prelimone}\sbox{\prelimone}{{\scriptsize\scriptbabar preliminary}}
\newsavebox{\prelimtwo}\sbox{\prelimtwo}{{\scriptsize\scriptbabar preliminary}}
\newsavebox{\prelimthr}\sbox{\prelimthr}{{\scriptsize\scriptbabar preliminary}}

\begin{document}

\thispagestyle{empty}

\hfill SLAC-PUB-10924 \\
\null\hfill hep-ex/0412065 \\
\null\hfill December 2004

\vspace{12mm}
\begin{center}
{\bf Radiative Penguin Decays of B Mesons:\footnote{Work supported by Department of 
Energy contract DE-AC02-76SF00515 and Department of Energy grant DE-FG05-91ER40622.} \\ 
Measurements of  $\B\to\Kstar\gamma$, $\B\to\Kstar_2(1430)\gamma$, and Search for $\Bz\to\phi\gamma$
}

\vspace{8mm}

Johannes M.~Bauer

\vspace{2mm}

Department of Physics and Astronomy \\
University of Mississippi--Oxford \\
University, Mississippi 38677, USA

\vspace{2mm}

{\sl representing the \BaBar Collaboration}

\vspace{2mm}

Stanford Linear Accelerator Center \\
Stanford University \\
Stanford, CA 94309, USA

\vspace{10mm}

{{\bf Abstract}
\vspace{3mm}

\begin{minipage}{5.2in}
      Electromagnetic radiative penguin decays of the
      $B$\ensuremath{\,}meson were studied with the \BaBar detector at
      SLAC's PEP-II asymmetric-energy \BF.  Branching
      fractions and isospin asymmetry of~the decay
      $B$$\rightarrow$$K^*\gamma$, branching fractions of
      $B$$\rightarrow$$K_2^*(1430)\gamma$, and a search for
      $B^0$$\rightarrow$$\phi\gamma$ are presented.  The decay rates may
      be enhanced by contributions from non-standard model processes.
\end{minipage}
}

\vfill

{\sl Presented at the 2004 Meeting of the Division of Particles and Fields \\
of the American Physical Society \\
Riverside, CA, USA \\
August 26, 2004 -- August 31, 2004 

\vspace{2mm}
Submitted to International Journal of Modern Physics A}
\vspace{6mm}
\null
\end{center}

\newpage

\section{Motivation and Data Analysis}

The decay of \B mesons into $\Kstar\g$ or $\Kstar_2(1430)\g$ is
forbidden at the tree-level, but allowed via one-loop $b\to s\g$
electromagnetic penguins~\cite{bib:penguin}.  Non-standard-model virtual
particles may take part in the loop and may affect the~decay rate.
Non-perturbative hadronic effects make theoretical predictions
difficult, but theoretical (as well as experimental) uncertainties are
reduced in ratios like the isospin asymmetry $\Delta_{0-}$, expected to
be 6 to 13\% in the Standard Model~(SM)~\cite{bib:isospin}:

\begin{equation}
\Delta_{0-} = \frac{ \Gamma(\Bzb\to\Kstarzb\g) - \Gamma(\Bm\to\Kstarm\g) }
                    { \Gamma(\Bzb\to\Kstarzb\g) + \Gamma(\Bm\to\Kstarm\g) }.
\label{eq:isospin}\end{equation}

Penguin annihilation dominates the very clean decay
$\Bz$$\to$$\phi\g$.  The branching fraction is only
$\BR(\Bz$$\to$$\phi\g) = 3.6\times10^{-12}$ in the SM, but higher with
R-parity violating super\-symmetry~\cite{bib:yang}.

The~results in this report originate from three independent analyses of
data collected by the \BaBar Detector~\cite{bib:nim} at the asymmetric
\BF, Stanford Linear Accelerator Center~(SLAC).  The~$\Kstar\g$ and the
$\Kstar_2(1430)\g$ analyses~\cite{bib:Kstarg,bib:Kstar1430g} use
$(\hbox{88 -- 89})\times10^6$ \BB events (82\invfb), while the $\phi\g$
analysis uses $124\times10^6$ \BB events (113\invfb).

The~hard photon selections are designed to especially remove daughters
of \piz or~$\eta$.  All kaons have to fulfill strict particle
identification criteria, and all \Kstar, $\Kstar_2(1430)$ or $\phi$
must satisfy additional requirements, \eg, on their mass.

The~major background comes from $\ep\en$$\to$$\qqbar$ decays
($q$$=$$u$$,$$d$$,$$s$$,$$c$).  Their jet-like structure is exploited to
distinguish them from the more spherically symmetric \BB events.  For
this, all~three analyses use neural networks with, \eg, variables based
on the directions or flavor of the particles in the event, as well as
helicity angles, which make use of correlations between
the direction of particles and their daughters.

The~final \B candidates are described by two variables, $\mes = \sqrt{
\rule{0pt}{8pt} E^{*2}_{\rm beam} - p^{*2}_B }$ and $\deltae = E^*_B -
E^*_{\rm beam}$, with $E^{*}_{\rm beam}$ the center-of-mass~(CM) energy
of the \ep/\en beam, and $E^*_B$ and $p^{*2}_B$ the CM energy and
momentum of the \B~candidate.

\section{Analysis of $\B\to\Kstar\g$ and $\B\to\Kstar_2(1430)\g$}

The following modes are reconstructed: 
   $\Bz$$\to$$\Kstarz\g$          with $\Kstarz$$\to$$\Kp\pim$            or $\KS\piz$, 
   $\Bp$$\to$$\Kstarp\g$          with $\Kstarp$$\to$$\Kp\piz$            or $\KS\pip$,
   $\Bz$$\to$$\Kstar_2(1430)^0\g$ with $\Kstar_2(1430)^0$$\to$$K^+\pim$, 
   $\Bp$$\to$$\Kstar_2(1430)^+\g$ with $\Kstar_2(1430)^+$$\to$$K^+\piz$  or $\KS\pip$.

The~number of signal events is extracted via maximum likelihood~(ML)
fits.  The signal shapes in \mes and \deltae are described by Gaussian and
Crystal Ball functions~\cite{bib:CB}, while the shapes of
\qqbar are described by ARGUS functions~\cite{bib:ARGUS} in \mes and 
first-order polynomials in~\deltae.  The~\BB background shapes are
determined from generic and exclusive Monte Carlo~modes.  The largest
contributor are other $\B$$\to$$X_s\g$ events.  For~$\Kstar_2(1430)$,
additional events come from $\Kstar(1410)\g$ and non-resonant
$\B$$\to$$\kaon\pi\g$ decays, which differ from signal in the
$\Kstar_2(1430)$ helicity angle~\thetaH.

\begin{figure}[t]
\centerline{\psfig{file=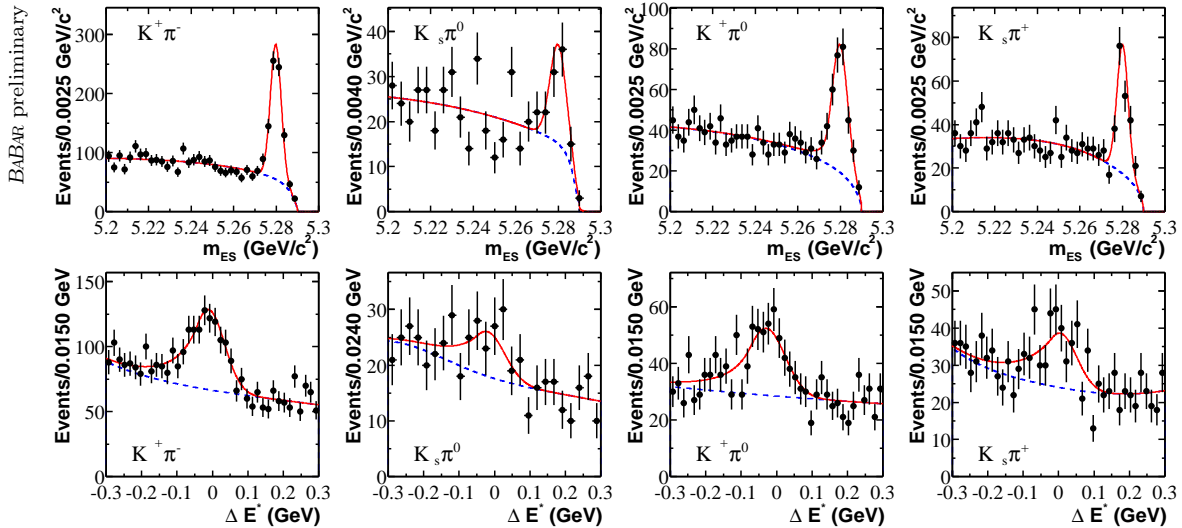,width=15cm}}
\vspace{-70mm}\hskip2mm\rotl\prelimone\vspace{44mm}
\caption{Distributions of \mes (top) and \deltae (bottom) for the four
modes in $\Kstar\g$.  Besides the data points, the full fit and the
background components are shown.\label{fig:MLF_Kstarg}}
\vspace{4mm}\null
\end{figure}

\begin{figure}[ht]
\centerline{\psfig{file=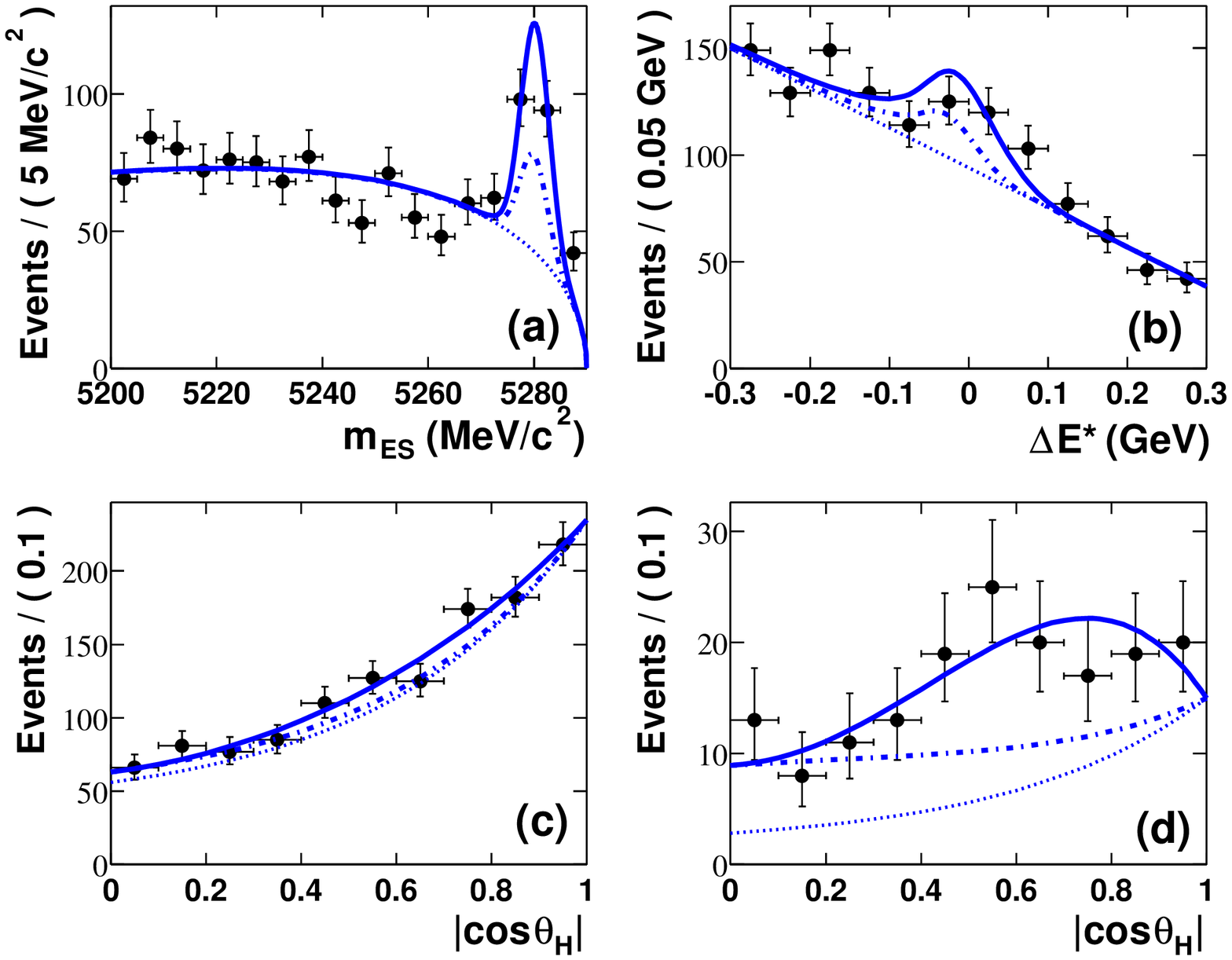,width=11.5cm}}
\vspace{-89mm}\hskip22mm\rotl\prelimtwo\vspace{61mm}
\caption{ML fit for $\Bz$$\to$$\Kstar_2(1430)^0\g$ and
$\Kstar_2(1430)^0$$\to$ $\Kp\pim$, with \mes (a), \deltae (b),
$\cos\thetaH$~(c), and $\cos\thetaH$ in the signal area
(d). The~points show data, the lines indicate peaking,
non-peaking, and signal contributions.
\label{fig:MLF_Kstar1430g}}
\vspace{4mm}\null

\end{figure}

The ML fits make use of \mes and \deltae (Fig.~\ref{fig:MLF_Kstarg}),
and for $\Kstar_2(1430)\g$ also of the $\Kstar_2(1430)$ helicity angle
(Fig.~\ref{fig:MLF_Kstar1430g}), and lead to the branching fractions
listed in Table~\ref{tab:BF_Kstarg}.  The isospin asymmetry $\Delta_{0-}
\hbox{(prelim.)} = 0.050 \pm 0.045\hbox{~(stat.)} \pm
0.028\hbox{~(syst.)} \pm 0.024\hbox{~$(R^{+/0}$,
Ref.~\refcite{bib:R+0})}$ is consistent with both the SM and previous
measurements.

\begin{table}[ht]
\tbl{Preliminary branching fractions (1$^{\rm st}$ error statistical, 2$^{\rm nd}$
     systematic).\label{tab:BF_Kstarg}}
 {\begin{tabular}{@{}ccccc@{}} \toprule
 mode                   & $\Bz$$\to$$\Kstarz\g$    & $\Bp$$\to$$\Kstarp\g$    & $\Bz$$\to$$\Kstarz_2(1430)\g$ & $\Bp$$\to$$\Kstarp_2(1430)\g$ \\ \colrule
 \BR~($\times 10^{-5}$) & $3.92 \pm 0.20 \pm 0.24$ & $3.87 \pm 0.28 \pm 0.26$ & $1.22 \pm 0.25 \pm 0.10$      & $1.45 \pm 0.40 \pm 0.15$      \\ \botrule
\end{tabular}}\end{table}

\section{Analysis of $\phi\g$}

\begin{figure}[ht]
\centerline{\psfig{file=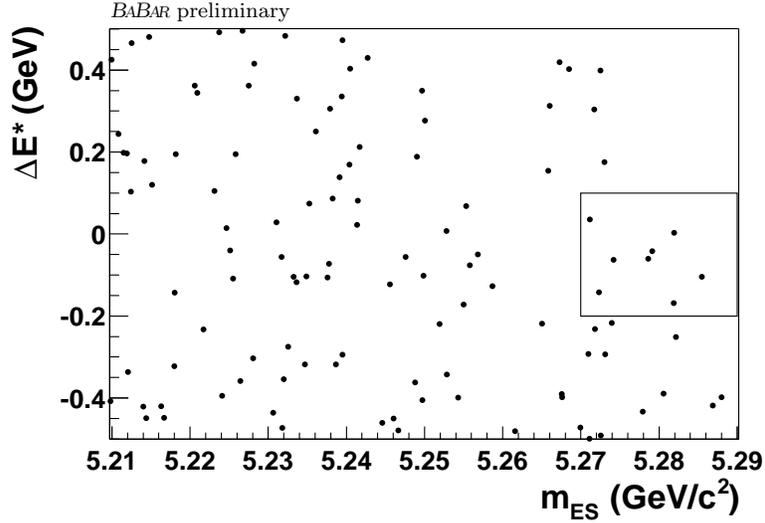,width=11cm}}
\vspace{-73mm}\hskip43mm\usebox\prelimthr\vspace{67mm}
\vspace*{-5pt}
\caption{Data candidates of $\Bz$$\to$$\phi\g$ in \mes-\deltae plane.
The~rectangle at the right side shows the limits of the signal
box.\label{fig:phig_2d}}\vspace{4mm}\null

\end{figure}

The mode $\Bz$$\to$$\phi\g$ is reconstructed with $\phi$$\to$$\Kp\Km$.
The~signal box is defined by $5.27 < \mes < 5.29\gevcc$ and $-0.2 <
\deltae < 0.1\gev$.  \BB background in the signal box is negligible
($0.09\pm0.05$ events from Monte Carlo).  Continuum background is estimated
to be $6\pm1$ events by fitting data events in the \mes and \deltae
regions outside the signal box and extrapolating into the signal box.
Since only eight events were found inside the signal box
(Fig.~\ref{fig:phig_2d}), the upper limit~\cite{bib:CH} for the branching
fraction of $\Bz\to\phi\g$ is $9.4\times10^{-7}$ at~90\%~confidence~level.

\clearpage

\section{Summary and Acknowledgment} 

All results of this report are preliminary.  The~branching fractions of
$\B\to\Kstar\gamma$ and $\B\to\Kstar_2(1430)\g$, as well as
$\Delta_{0-}$ were measured and are in agreement with previous
measurements and SM predictions.  The~upper limit on the branching
fraction of $\Bz$$\to$$\phi\g$ is currently the tightest published limit
on this mode.  The~lack of signal is consistent with the Standard~Model.
 
The~author thanks the \BaBar collaboration, the SLAC accelerator group
and all contributing computing organizations.  He~was supported by
U.S. Dept.~of Energy grant DE-FG05-91ER40622.

\end{document}